\documentclass[pre,letterpaper]{revtex4}
\usepackage[dvips]{graphicx}
\setkeys{Gin}{width=0.4\columnwidth}
\usepackage{amsmath}

\newcommand{\dparcial}[2]{\frac{\partial\,#1}{\partial#2}}
\newcommand{\ddparcial}[2]{\frac{\partial^2\,#1}{\partial#2^2}}

\begin{document}
\title{Study of Arbitrary Nonlinearities in Convective Population Dynamics
with Small Diffusion}
\author{I. D. Peixoto} 
\author{L. Giuggioli}
\author{V. M. Kenkre} 
\affiliation{Consortium of the Americas
      for Interdisciplinary Science and Department of Physics and
      Astronomy, \\ University of New Mexico, Albuquerque, New
      Mexico 87131}


\begin{abstract}
  Convective counterparts of variants of the nonlinear Fisher equation
  which describes reaction diffusion systems in population dynamics
  are studied with the help of an analytic prescription and shown to
  lead to interesting consequences for the evolution of
  population densities. The initial value problem is solved explicitly
  for some cases and for others it is shown how to find traveling wave
  solutions analytically. The effect of adding diffusion to the convective 
  equations is first studied through exact analysis through a piecewise 
  linear representation of the nonlinearity. Using an appropriate small 
  parameter suggested by that analysis, a perturbative treatment is developed 
  to treat the case in which the convective evolution is augmented 
  by a small amount of diffusion.
\end{abstract}

\date{\today} \maketitle

\section{Introduction}
The Fisher equation, proposed originally to describe the spread of an
advantageous gene in a population \cite{fisher1937}, has found a
great deal of use in mathematical ecology, directly
\cite{murray,skellam,kk,bkk} as well as indirectly,
i.e., in modified forms. Examples of modifications are the
incorporation of internal states such as those signifying the presence
or absence of infection \cite{ak,akyp,kpasi}, 
the
introduction of temporal nonlocality in the diffusive terms
\cite{manne,abk}, of spatial nonlocality in the competition
interaction terms \cite{fkk,fk}, and the addition of convective terms
signifying `wind effects' \cite{nelson,ana,gk,lgthesis,k}.  Most of
the modifications lead to equations which, as in the case of the
original Fisher equation, are not soluble analytically and require
approximate or numerical methods for their analysis. There is,
however, one modification \cite{gk}, that leads to an analytic
solution via a trivial transformation, and to the possibility of the
extraction of interesting information of potential use to topics such
as bacterial population dynamics. That modification consists of the
replacement of the diffusive term in the Fisher equation
\begin{equation} \label{eq:fisher}
\dparcial{u}{t}= a u \left(1-\frac{u}{K}\right) + D \ddparcial{u}{x}
\end{equation}
by a convective term. As is well-known, in the original Fisher
equation, $u$ represents a concentration or density, $a$ the growth
rate, $K$ the so called `carrying capacity', and $D$ the diffusion
constant. In spite of the fact that purely convective (non-diffusive)
nonlinear partial differential equations are much less sophisticated
than their diffusive counterparts, there are at least two reasons why
their study is important. There are clear physical situations
\cite{nelson,ana,gk} in which wind effects which add the convective
term to a diffusive equation are present.  These arise, for instance,
when one changes one's reference frame to that of a moving mask in
bacterial population dynamics experiments \cite{nelson}. Because the
convective term comes from the macroscopic motion of the mask under
experimental control whereas the diffusive term comes from the
microscopic motion of the bacteria, it is possible to arrange the
system experimentally so that the diffusion effects are relatively
unimportant. This may be achieved for instance, by using viscous
environments, or by genetically engineering the (bacterial)
population. In such a case one arrives at the analysis of a largely
convective (negligibly diffusive) nonlinear evolution which may be
investigated in zeroth order as a purely convective equation
\cite{gk,k}. This fact, that experimental realization of the
convective nonlinear equation is indeed possible, provides one
motivation for these studies. The other, as explained in Ref.
\cite{gk}, is that it is possible to develop perturbation schemes
starting from the soluble convective equation to incorporate diffusive
effects. In Sections \ref{sec:application} and \ref{sec:travelfront},
we develop the theory for arbitrary nonlinearities in the absence of
the diffusion constant, and in Section \ref{sec:diffusion}, 
we focus on the effect of
adding the diffusive term.

Recently, a simple prescription for generalizing the analysis of Ref.
\cite{gk} was given by one of the present authors \cite{k} to
arbitrary nonlinearities in the convective equation. The equation
considered there is
\begin{equation} \label{problem}
\dparcial{u}{t} + v \dparcial{u}{x} = auf(u),
\end{equation}
where $v$ is the medium or `wind' speed, $a$ is a growth rate and
$f(u)$ is a dimensionless \emph{arbitrary} nonlinearity. The
prescription provided in that analysis gives the full time and space
evolution of $u(x,t)$ in terms of its initial distribution
$u(x,0)=u_0(x)$, through the relation
\begin{equation} \label{solution}
 u(x,t) = G^{-1} \big[G[u_0(x - v t)] - at\big]
\end{equation}
where the function $G(u)$ is obtained from the nonlinearity $f(u)$ by
the integration \cite{k}
\begin{equation} \label{eq:gdef}
G(u) = - \int \frac{1}{uf(u)} du.
\end{equation}
We have used a slightly changed notation here relative to ref.
\cite{k}.  Our functions $G(u)$ and $f(u)$ are both dimensionless in
contrast to the dimensioned counterparts $g(u)$ and $F(u)$ in ref.
\cite{k}.

In this paper we present several applications of the prescription
(\ref{solution}) and report a number of additional results from the
nonlinear convective equation (\ref{problem}). In Section \ref{sec:application}, we
examine features of the solutions obtained exactly from the
prescription for several cases of the nonlinearity including an
interesting `pyramid effect' that occurs for nonlinearity functions
$f(u)$ with \emph{multiple zeros}.  In Section \ref{sec:travelfront}, we develop a general
analysis for finding traveling wave solutions, and to study
characteristics of the tails and shoulders of such solutions.
Applications of this analysis for the nonlinearities studied in
Section \ref{sec:application} are also made in Section \ref{sec:travelfront}.  
In Section \ref{sec:diffusion}, we analyze in
detail the perturbative problem obtained by adding a small amount
of diffusion to the nonlinear convective equation.  
Conclusions are presented in Section \ref{sec:conclusions}.

\section{ Applications of the Prescription for the Solution of the
  Initial Value Problem} 
\label{sec:application}
Applications of the prescription
(\ref{solution}) for the solution of the arbitrary initial value
problem were indicated in ref. \cite{k} for three cases: the ordinary
logistic nonlinearity for which $f(u)$ has a single zero at $u=K$, the
Nagumo nonlinearity for which $f(u)$ has two zeros, and the
trigonometric nonlinearity for which $f(u)$ has multiple zeros. We
will display new results for the latter two, and analyze other
generalizations of the logistic nonlinearity. In some of these cases  (subsections A,D and E below) 
$u(x,t)$ is presented in an implicit form, whereas in the others 
(subsections B and C) the form 
is given explicitly.

\subsection{Allee effect from Nagumo nonlinearity}
\begin{figure}[!htbp]\includegraphics[width=\columnwidth]{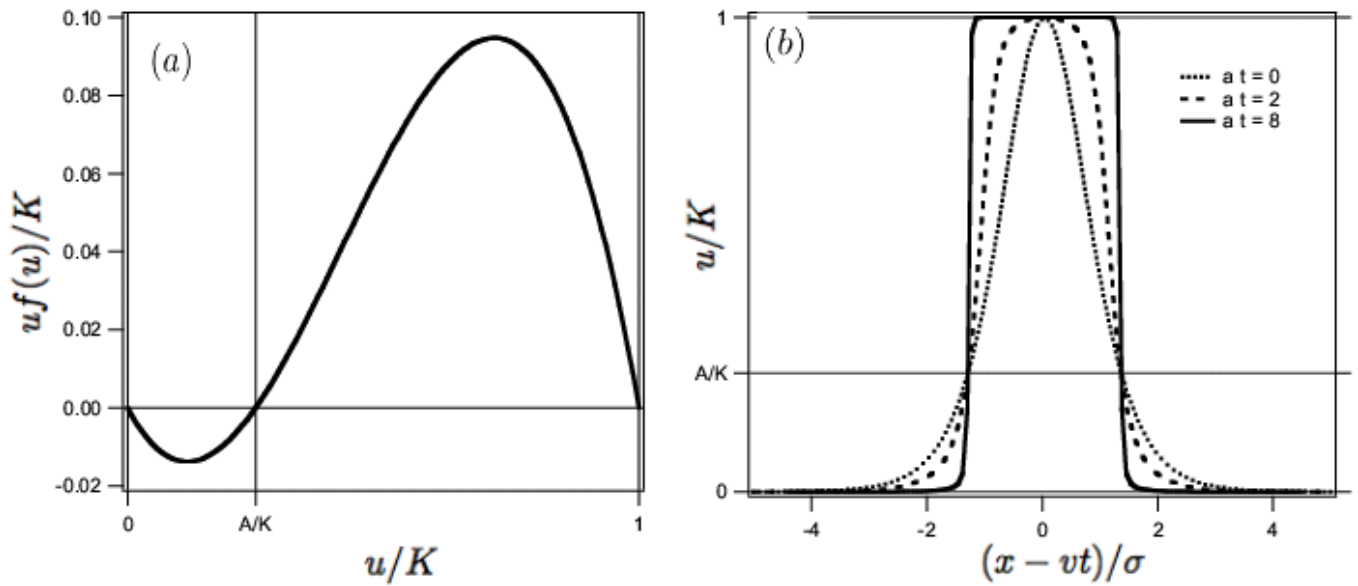}
\caption{The Nagumo nonlinearity of Eq.~(\ref{nagumof}) and its effects
  on evolution. The nonlinearity is plotted in (a). Shown in (b) is an
  example of the evolution from an initial condition $\textrm{sech}(x/\sigma)$
  (shown by the dotted line) whose profile belongs both to the region
  where $0<u(x)<A$ and where $A<u(x)<K$.  The portion in the first
  (lower) region decays while that in the second (upper) region grows
  to the saturation value $u(x)=K$.  Here, $\sigma$ is an arbitrary
  length.  }\label{fig:nagumo}
\end{figure}

In (\ref{problem}) let $f(u)$ be given by
\begin{equation}\label{nagumof}
f(u)= \left(1 - \frac{u}{K}\right)
\left(\frac{u}{K}-\frac{A}{K}\right)
\end{equation}
where $A<K$. Such an $f(u)$ represents the so called `Allee effect'.
This effect \cite{allee} is associated with the existence of an
additional zero (fixed point) in the nonlinearity relative to the
logistic case. Unlike in the logistic case, the zero-$u$ solution is
stable here. If $u$ is small initially, it is attracted to the
vanishing value; if large, it is attracted to the non-zero value
$u=K$.  The demarcation point is the additional zero introduced in
this case, viz. $u=A$.  The physical origin of the Allee effect in population
dynamics is the possible increase of survival fitness as a function of population size for 
low values of the latter. Existence of other members of the species may induce 
individuals to live longer whereas low densities may, through loneliness, lead to extinction.
There is ample evidence  for such an effeect in 
nature \cite{courchamp, SS, fowlerbaker}.To construct $G(u)$ from (\ref{eq:gdef}) is
straightforward,
\begin{equation}
G(u)=\frac{1}{A(K-A)}
\ln\left(\frac{(K-u)^{A}u^{K-A}}{|u-A|^{K}}\right),
\end{equation}
but to invert $G(u)$ is not.  In Fig.~(\ref{fig:nagumo}) the
nonlinearity is displayed along with the time evolution for a certain
initial condition. The Allee effect causes a decay to zero of parts of
the solution, while other parts evolve to the saturation value K.
Thus, an initial condition with $u$ values belonging to both regions,
will show a growth for those points in space such that $u_{0}(x)>A$,
while those points such that $u_{0}(x)<A$ will decay to zero.  The
initial condition in this case evolves into a step profile of width
given by the separation in space of the two points where $u_{0}(x)=A$.

\subsection{Trigonometric nonlinearity and the Pyramid Effect}
\begin{figure}[!htbp]
  \includegraphics[width=\columnwidth]{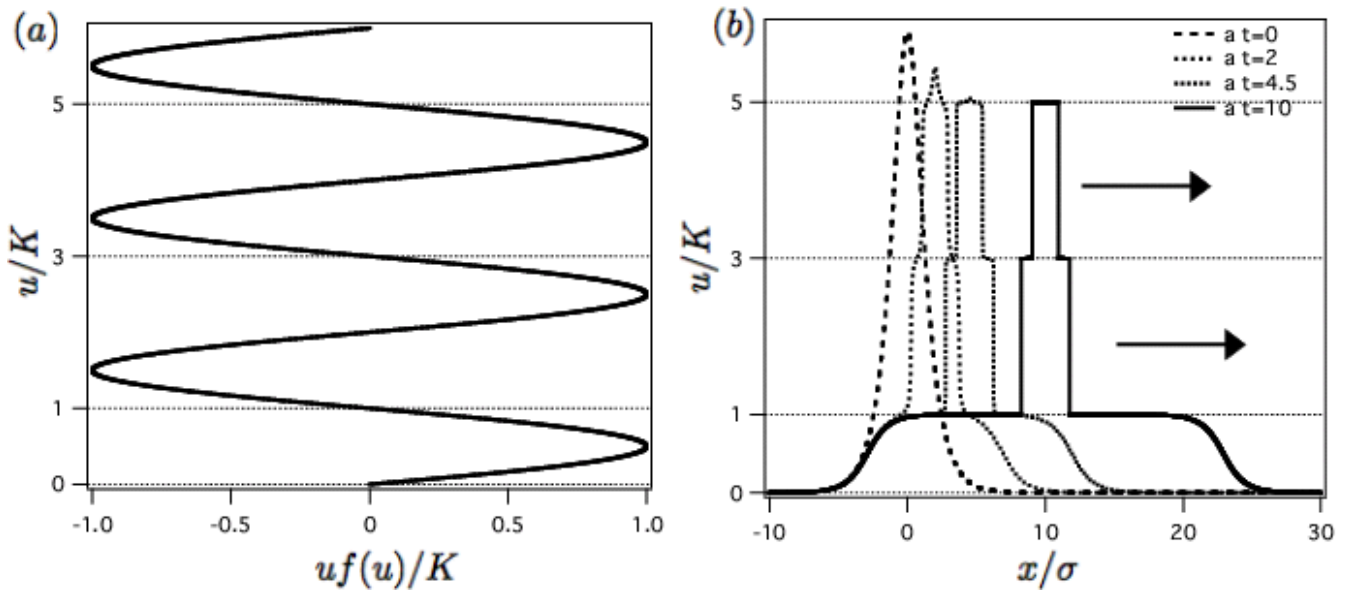}
\caption{The `pyramid effect' for a nonlinearity with multiple zeros. The sinusoidal nonlinearity of 
  Eq.~(\ref{nlpyramid}) is plotted in (a). The horizontal lines at
  $u/K=1,3,5$ depict levels of separation of the regions
  between stable zones. The corresponding evolution is plotted in (b)
  for the initial profile $u_{0}(x)=5.9\,K\,\textrm{sech}(x/\sigma)$, where
  $\sigma$ is an arbitrary length. The initial profile is shown as a
  dotted line. The profile moved rightwards as
  shown.}\label{fig:sinusoidal}
\end{figure}

What would happen if the nonlinearity function $uf(u)$ had multiple zeros
rather than three as in the Nagumo case? The physical origin of such
behavior may lie in nonmonotonic response of survival fitness to population size.
Complex biological and sociological interactions among members of the species 
could exist on the basis of the coupling of mating instincts with mere size effects.
Such interactions could, in principle, make the survival fitness optimum at several, rather than a
single, values of the population size. In order to address such situations, extensions of the normal
Allee considerations are necessary. For this purpose, the discussion in A above may be generalized as follows. For a given
nonlinearity with an arbitrary number of zeros, $uf(u)$ can be
sectioned among each subsequent unstable zero. The dynamics of each
point of the initial profile will evolves according to which of these
sections it belongs initially. In the Nagumo case there are only two
such regions: the first region is between $u=0$ and $u=A$, while the
second region is for $u>A$. All the points of the initial condition
such that $u_{0}(x)<A$ decay to zero, while all those points belonging
to the second region grow to the saturation value $u=K$. But when
there is more than one unstable zeros, as in the case of a
trigonometric nonlinearity, the propagating solution may evolve into a
pyramidal profile. To show this we choose a sinusoidal nonlinearity
given by
\begin{equation} \label{nlpyramid}
f(u)=\frac{K}{\pi }\frac{\sin \left( \frac{\pi u}{K}\right) }{u}
\end{equation}
where we took the opportunity to change the stability of $u=0$ compared with the Nagumo case presented in A.
From (\ref{eq:gdef}), $G(u)$ is given by
\begin{equation}
G(u)=-\ln \left[ \tan \left( \frac{\pi u}{2K}\right) \right] 
\end{equation}
and the exact solution at all time is
\begin{eqnarray}
u(x,t) &=&\sum_{m=1}^{+\infty }K\left( 1-\frac{2}{\pi }\arctan \left[
e^{-at}\cot \left( \frac{\pi }{2K}\left\{ \Theta \left[ \frac{2mK}{\pi }%
-u_{0}\left( x-vt\right) \right] \right. \right. \right. \right.   \nonumber
\\
&&\left. \left. -\Theta \left[ \frac{\left( 2m-2\right) K}{\pi }-u_{0}\left(
x-vt\right) \right] u_{0}\left( x-vt\right) \right) \right] +\left(
m-1\right) \times   \nonumber \\
&&\left. \left\{ \Theta \left[ \frac{2mK}{\pi }-u_{0}\left( x-vt\right)
\right] -\Theta \left[ \frac{\left( 2m-2\right) K}{\pi }-u_{0}\left(
x-vt\right) \right] \right\} \right) .
\end{eqnarray}
Let us consider an initial condition such that $0<u_{0}(x)<2mK$ where
$m$ is a positive integer. All those initial points of the profile
that belong to the region delimited by $(2m-2)<u(x)<2mK$ can be
described by the following dynamics. The solution grows from
$u=(2m-2)K$ to the value $u=(2m-1)K$ and decays to it from $u=2mK$
following the evolution of an initial condition with compact support
in that region. As extensively studied in the case of the logistic
nonlinearity \cite{gk}, any initial condition with a bounded domain
evolves into a step profile of width given by its initial value. In the
sinusoidal nonlinearity, from each region a step profile with width given by
 $%
\Theta \left[ 2mK/\pi -u_{0}(x)\right] -\Theta \left[ (2m-2)K/\pi %
  -u_{0}(x)\right] $ eventually emerges. Figure (\ref{fig:sinusoidal})
shows an example for an initial condition that overlays three such
regions. It is clear from the figure that in the bottom region the
profile behaves as if the other regions did not exist.

\subsection{Multi-individual struggle for environment resources: linear growth with power saturation}
\begin{figure}[!htbp] 
\begin{center}
  \includegraphics[width=\columnwidth]{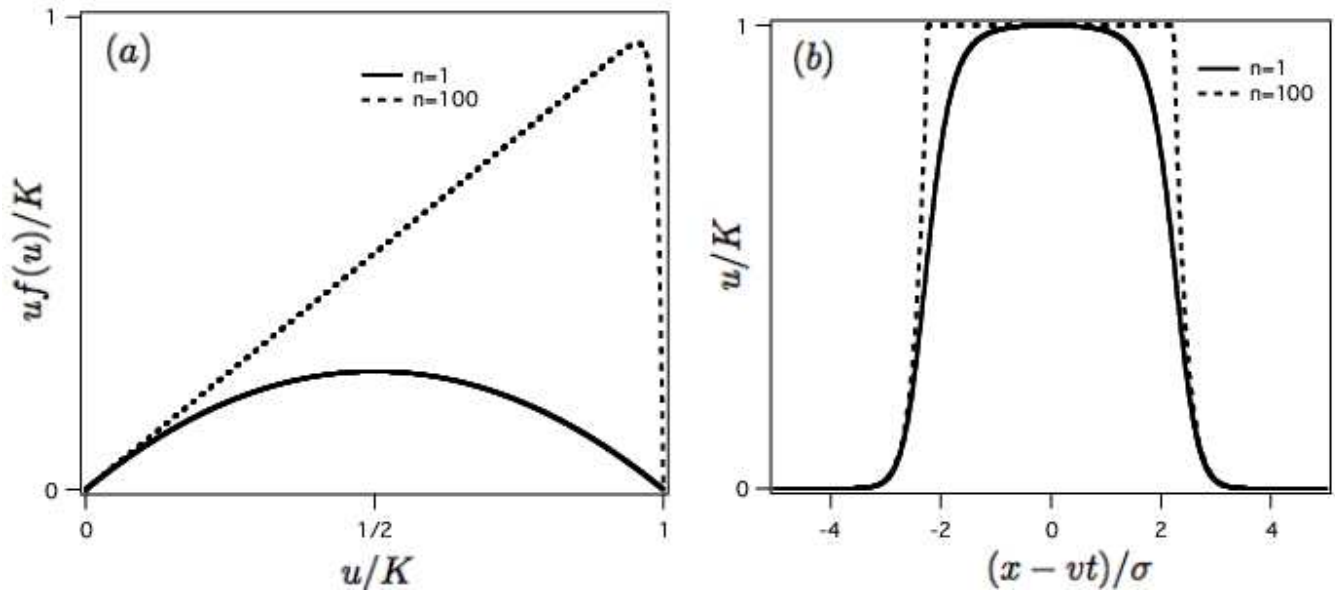}
\caption{The nonlinearity with
  power saturation and its effect on the time evolution.  The
  nonlinearity with $f(u)$ given by (\ref{eq:lgps}) is shown in (a) for two different
  powers $n$.  Near $u=0$ both functions are similar but for higher
  values they differ dramatically since the case $n=100$ has an abrupt drop close
  to $u=K$. Their corresponding time evolution is plotted in
  (b) at a time $at=5$ starting from the same initial condition $K \textrm{sech}(x/\sigma)$
  where $\sigma$ is an arbitrary length.}
\label{fig:pictorialgenlog} 
\end{center} 
\end{figure}

One can envisage physical situations in which one encounters a simple generalization of the logistic nonlinearity that retains
the linear behavior near $u=0$ but accentuates the nonlinearity for
larger values of $u$. This would arise, e.g., if the elemental struggle of the individuals for resources involves
not two but multiple individuals. The struggle would then not be binary in nature. To address this  possibility, we consider
\begin{equation} \label{eq:lgps} 
f(u)= \left[ 1 - (u/K)^{n}\right].  
\end{equation} 
The saturation effect already seen in the logistic case, becomes more
abrupt as $n$ increases.  Pictorial representations for several cases
of $n$ are provided in Fig.~(\ref{fig:pictorialgenlog}).  Application
of the prescription (\ref{eq:gdef}) produces the explicit form for
$G(u)$
\begin{equation} G(u)
=\frac{1}{n} \ln \left[ \frac{1 - (u/K)^n}{(u/K)^n} \right].  
\end{equation}
Its inverse can be written analytically. An arbitrary initial
condition $u_0 (x)=u(x,0)$ therefore via prescription (\ref{solution})
evolves according to
\begin{equation} u(x,t) =
\frac{u_0(x-vt)}{\left[\left(\frac{u_0(x-vt)}{K}\right)^n(1-e^{-ant})+e^{-ant}\right]^{1/n}}.
\end{equation} 

\subsection{Nonlinear growth with power saturation as in bisexual reproduction}
Interesting consequences arise if the nonlinearity cannot be
approximated by a linear term near $u=0$, such as when the growth of
the species is nonlinear. A physically relevant example is the
representation of bisexual reproduction that would require a growth
term bilinear rather than linear (see for example Ref.~\cite{rosas2002}).
The generalization of the nonlinearity (\ref{eq:lgps}) to a
\emph{nonlinear growth} term can be expressed as
\begin{equation} \label{eq:nonlinearf}
f(u)= (u/K)^{m}\left[1-(u/K)^{n}\right]
\end{equation}
where $m+1$ is the first power in $u$ in a Taylor expansion near zero.
Prescription (\ref{solution}) gives
\begin{equation} \label{eq:nonlinearG}
G(u)=\frac{{}_{2}F_{1}\left[1,-\frac{m}{n},1-\frac{m}{n},
\left(\frac{u}{K}\right)^{n}\right]}%
{K m \left(\frac{u}{K}\right)^{m}}.
\end{equation}
where ${}_{2}F_{1}$ is the hypergeometric function.  In general, Eq.
(\ref{eq:nonlinearG}) cannot be inverted and $G^{-1}$ has to be
calculated implicitly.  However, analytical considerations regarding
the shape of tails and shoulders of traveling fronts can be easily
obtained through the help of Eq. (\ref{eq:nonlinearG}) as will be seen
below in Sec \ref{sec:travelfront}.

\subsection{Nonlinearity with provisional saturation}
\begin{figure}[!htbp]
\begin{center}
  \includegraphics[width=\columnwidth]{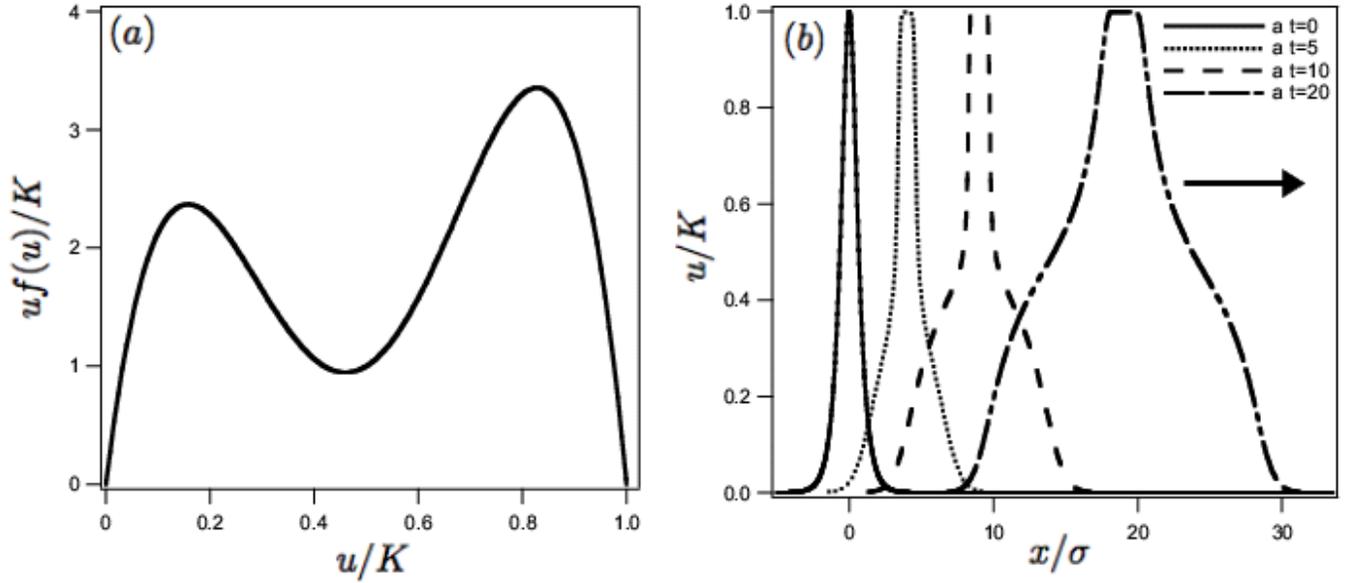}
  \caption{Example of a nonlinearity with provisional saturation. 
  In (a) the reaction function $uf(u)/K$ of the polynomial
  \ref{eq:poly1} is presented, while in (b) the corresponding evolution
for an initial condition is shown.}
\label{fig:dip}
\end{center}
\end{figure}

\begin{figure}[!htbp]
\begin{center}
  \includegraphics[width=\columnwidth]{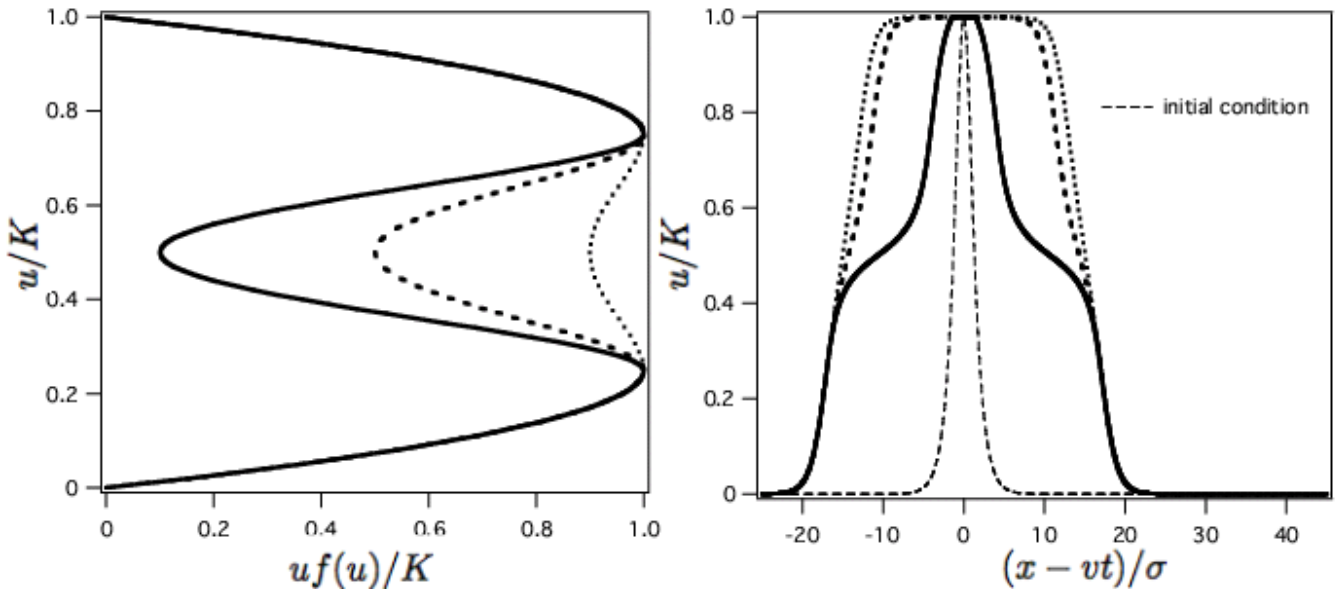}
\caption{Dependence of the depth of the dip for nonlinearities with provisional saturation. In (a) three different nonlinearities are plotted. The axis are interchanged respect to the previous figure.
This has been done to express clearly the correspondence between the nonlinearities in (a) and the evolution in (b).}
\label{fig:dip2}
\end{center}
\end{figure}

The previous nonlinearities we have discussed above increase
(decrease) monotonically from one zero to a maximum (minimum) and
then monotonically decrease (increase) to the next zero.  Our
prescription (\ref{solution}) permits the analysis of more interesting
situations that may arise if the mutual struggle for resources among individuals constituting
the species leads to saturation of the population size that is not permanent but only provisional.
Further increase in the population density in such a situation might lead to a returned
increase in the nonlinearity, followed by an eventual decrease. Such a state of affairs
is analogous to the case analyzed in subsection B above but with an important difference.
In the present case, $uf(u)$ does not dip below zero. As a consequence, there
are no multiple zeros at $u \neq 0$. In Fig.(\ref{fig:dip}) we present, as an example of such a
nonlinearity, a polynomial of the form,
\begin{equation} \label{eq:poly1}
u f(u)=9(u/K) - \frac{131\,(u/K)^2}{12} + \frac{9\,(u/K)^3}{2} - 
  \frac{7\,(u/K)^4}{12}
\end{equation}
and the corresponding time evolution.  As time
evolves, the curvature of the solution reflects the specific
particulars of the ``reaction'' term. 
In the solution the regions with $u$ values close to the top of
the reaction function have more inclination.
To study the effects of different dips, in Fig.(\ref{fig:dip2}) three similar $u f(u)$ are shown 
which differ in the 'depth' of their valleys. Their construction is straightforward by means of the Lagrange interpolating polynomial, i.e. by taking
a polynomial of fourth order and requiring it to have specific values at the five points desired.
In the example of the figure, four of the five points that are required to be common are: $(0;0),(1;0),(1/4;1),(3/4;1)$
described in $(u/K;u f(u))$ plane. The other point will characterize the dip, and
for the example shown in the figures are $(1/2;9/10),(1/2;1/2),(1/2;1/10)$.
From the same initial condition,
the evolution will differ as shown in (b): the more pronounced
the dip in the nonlinearity, the flatter the solution close to the
center of the dip.

\section{Traveling Front Solutions: Analytical expressions for spatial limits} 
\label{sec:travelfront}
The analysis in the previous section has focused on the initial value
problem. Much interesting information in nonlinear problems may
be extracted, however, directly from traveling waves (see, e.g., extensive
reports in the soliton literature such as in \cite{soli}). The $G$-function appearing in our prescription
(\ref{solution}) can be used conveniently, and directly, to find
traveling wave solutions. As will be seen below, there are situations
in which this technique is useful even when explicit analytic
expressions cannot be found for the initial value problem. Imposing
the traveling wave ansatz $u(x,t) = U( z = x - c t)$ in
Eq.\eqref{problem},
\begin{equation} \label{eq:tfronteq}
-\beta \dparcial{U}{z}= a U f(U),
\end{equation}
where 
\begin{equation}
\beta=c-v
\end{equation}
is the difference between the traveling front velocity
and the medium velocity.
Integrating (\ref{eq:tfronteq}) in $z$, leads to the solution
\begin{equation}\label{eq:tweq}
G\left[ U_{\beta }(z)\right] =a\frac{z}{\beta },
\end{equation}
where $G$ is defined in Eq.~(\ref{eq:gdef}).      
If the function $G^{-1}$ can be written analytically, the
front solution can be obtained in explicit form:
\begin{equation}
U_{\beta }(z)=G^{-1}\left[ a\frac{z}{\beta }\right] .
\label{eq:tfteo}
\end{equation}
Otherwise it is defined implicitly through Eq. (\ref{eq:tweq}).

\begin{figure}[!htbp]
\begin{center}  \includegraphics[height=0.4\columnwidth,width=0.45\columnwidth]{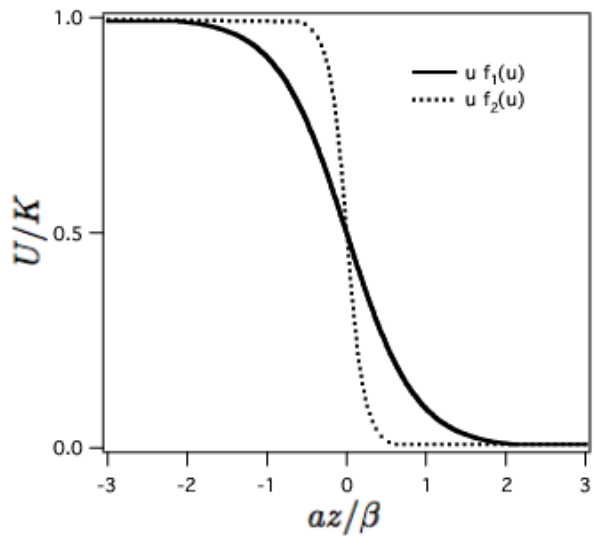}
\end{center}
\caption{Example of several traveling waves of the \emph{same speed, $\beta =1$}, but
  for different nonlinearity functions $f$. The solid line is the
  generalized logistic function, $u f_1(u)=u(1-u^{2})$, while the dotted
  line comes from the first half period of the sinusoidal function, $u f_2(u)=sin(\pi
  u)$. }
\label{fig:tw2speed1}
\end{figure}

Equation (\ref{eq:tfteo}) shows that, for a given nonlinearity, the
steepness of the front depends on the propagation velocity $c$ while
its shape is determined by the function $G^{-1}$. In order to show
how, for a given propagation velocity $c$, the shape of the front
changes for various nonlinearities (according to the simple relation
(\ref{eq:tfteo})), we provide Fig.  (\ref{fig:tw2speed1}).

We now show how one may find analytic expressions for the tail and the
shoulder of the traveling wave through inspection of the limiting
values of the nonlinearity $auf(u)$, respectively, as $u\to
0^{+}$ and $u\to K^{-}$, where $0$ and $K$ are both zeros of
the nonlinearity. The basic idea is to apply to Eq. (\ref{eq:tweq}) a
judiciously chosen function $h$, multiply by $U$ or $(K-U)$ as
appropriate, and then calculate the relevant limit of the product.
Thus, if we are interested in the spatial dependence of the tail of
the solution, we seek to obtain
\begin{equation} \label{taillimit}
\lim_{U\to 0^{+}} U(z) h_T\left( G\left(
U\right)
\right) =\lim_{z/\beta \to +\infty }  U(z) h_T(az/\beta ) = C_T.
\end{equation}

The function $h_T$ is chosen in such a way that the limit for
$U\to 0^{+}$ of the left hand side of the equation equals a
non zero constant depicted above by $C_T$.  If the shoulder spatial
dependence is sought, we apply another function $h_S$, and multiply by
$(K-U)$ rather than by $U$ before taking the limit $U\to
K^{-}$, 
the choice of $h_S$ being such that the limit is a constant
$C_S$.

The spatial dependence of the traveling solution in the two extreme
limits is found to be
\begin{equation} \label{eq:taillimitsolution}
\lim_{z/\beta \to + \infty }U(z)\simeq \frac{C_{T}}{h_{T}\left(
az/\beta\right) } 
\end{equation}
for the tail and
\begin{equation}
\lim_{z/\beta \to - \infty }U(z)\simeq K-\frac{C_{S}}{h_{S}\left(
az/\beta\right) }  \label{shouldep}
\end{equation}
for the shoulder.

The steepness of the front at the inflection point is another
important feature of the traveling front that can be obtained from the
shape of the nonlinearity even if the exact analytic form is not
known. From Eq. (\ref{eq:tfronteq}) we can simply relate
\begin{equation}
\frac{dU(z)}{dz}=-aU(z) \frac{f\left[ U(z)\right] }{\beta}.
\end{equation}
which clearly shows the relation between the maximum of $f(u)$ and the
front derivative at the inflection point. Comparing such values with
those cases of $f(u)$ in which the traveling front shape are known
analytically, allows to determine further qualitative characteristics
of the front shapes. We apply these various general considerations to
two specific cases below.

\subsection{Linear growth with power saturation}
From the general expression (\ref{solution}), it is straightforward to
write the analytic form of the traveling fronts
\begin{equation}
U_{\beta}(z) = \frac{K}{\left(1 + \left(2^{n}-1\right)e^{an
\frac{z}{\beta}}\right)^{1/n}},
\label{travfrontsat}
\end{equation}
where we have chosen to center the traveling front in such a way that
$U_{\beta}(z=0)$ is half the saturation value of $U$.
Application of the limiting procedure described above shows that
regardless of the value of the power saturation $n$ the front tail is
given by
\begin{equation}
\lim_{z/ \beta\to +\infty}U_{\beta}(z) \propto e^{-a z/\beta}.
\end{equation}
The value $n$ does not play a role since the tail shape is determined
by the form of the nonlinearity close to $U=0$ where saturation
effects are negligible. On the other hand, the value $n$ does play a
role in determining the shoulder shape.  The limiting procedure gives
in fact
\begin{equation}
\lim_{z/ \beta\to -\infty}U_{\beta}(z) \propto K-\frac{K}{n}e^{-a z/\beta}.
\end{equation}
which coincides with the one calculated directly from Eq.
(\ref{travfrontsat}).

It is interesting to notice that in the limiting case corresponding to
$n\to \infty$ the traveling front solution can be written as
\begin{equation}
U_\beta(z)=\left\{\begin{array}{ll}
K & \textrm{if } z\leq 0\\
K e^{-z/\beta} & \textrm{if } z>0
\end{array}\right.
\end{equation}

\subsection{Nonlinear growth}
For the case in which the growth near $u\sim 0$ is bilinear as in
(\ref{eq:nonlinearf}) for $m=2$, the prescription (\ref{solution}) is given in terms of
hypergeometric functions (see Ref.~\cite{hypergeometric}), whose inverses are not
known. However our limit prescriptions (\ref{eq:taillimitsolution}) and
(\ref{shouldep}) do provide analytical expressions for the spatial
limits.  For $U \approx 0$ the following behavior is obtained
\begin{equation}
G(u) \propto \frac{1}{K m (U/K)^m}.
\end{equation}
The function $h_T$ of Eq.\eqref{taillimit} which makes the left limit
a constant is found to be
\begin{equation}
h_T(G)= G^{\frac{1}{m-1}}.
\end{equation}
Therefore, in the tail,
\begin{equation}
\lim_{z/\beta \to +\infty} U(z) \simeq \frac{C}{(z/\beta)^{\frac{1}{m-1}}}.
\end{equation}

Indeed, we can determine the shape of the tail (shoulder) for any function $f$
that has a Taylor expansion near $U=0$ ($U=K$).

\section{Incorporation of diffusion along with convection}
\label{sec:diffusion}
The analytic treatment we have given above of the reaction-convection
problem represented by Eq. (\ref{problem}) has been made possible by
the fact that the partial differential equation considered is of first
order. Once the transformation is made to convert (\ref{problem})
exactly into its linear counterpart, the method of characteristics \cite{char},
well-known in the context of linear equations, is what is
behind the analysis we have presented.
If diffusion
is added to (\ref{problem}), individual points of the $u(x)$ curve at
a given time no longer evolve independently as they do for systems
characterized by (\ref{problem}) but influence one another in the
evolution.  Finding analytic solutions in the manner
detailed here and in Refs. \cite{gk,k} is then impossible. There is no
doubt from the practical viewpoint that the case $D=0$ is important to
study. This is so because negligible diffusion does occur in several
physical
situations as in the case of bacterial dynamics when the environment
or genetic engineering can be made to result in convection being far
more important than diffusion. Nevertheless it is important to ask
whether the introduction of diffusion can be considered a perturbation
or whether it alters the problem drastically, even when it is
relatively small in magnitude. If the latter were true, the analysis
presented in the previous sections would be uninteresting for all
realistic cases in which
finite diffusion exists. We therefore analyze the effects of adding
diffusion to the nonlinear convective problem:
\begin{equation}
  \label{eq:driftdiffusion}
  \dparcial{u}{t} + v \dparcial{u}{x} = a u f(u) + D \ddparcial{u}{x}.
\end{equation}
We do this in three parts, always focusing attention on traveling wave
solutions for simplicity. In the first part (subsection A below), we
obtain exact analytical expressions
for traveling wave solutions of the reaction diffusion problem
represented by equation (\ref{eq:driftdiffusion}) when $f$ is a
piecewise
linear function.  Also,  we compare the results between the absence of
diffusion to those in its presence (arbitrary amount of diffusion). In
the second part, when the amount of diffusion is small, we identify an
appropriate small parameter and develop
a perturbation scheme. In the third part, we apply the
perturbation treatment to one of the interesting nonlinearities
(sinusoidal) introduced at the beginning of this paper. Furthermore,
we analyze its validity by comparing it to the exact solution of (\ref{eq:driftdiffusion}) when
the nonlinearity is piecewise linear. The last two parts make up subsection B below.

\subsection{Diffusion effects treated through a piecewise linear representation of the nonlinearity}

Although \emph{exact} analytic solutions of the nonlinear equation we consider are
impossible in the presence of diffusion, they are possible for
traveling waves if we employ a piecewise linearization of the
nonlinearity. The piecewise linear approximation has been used
earlier in the analysis of the Fisher equation for the study of the effects of
transport memory \cite{manne} as well as pattern formation
\cite{horacio}. In both cases it was possible to extract useful
information through this resource, whose advantage is that once
the approximation of representing the given nonlinearity
by the piecewise linear form is made, no further approximation is invoked.
Let us begin with a piecewise representation of the logistic nonlinearity through
\begin{equation} \label{eq:linear}
f(u)=
\left\{\begin{array}{ll}
1 & u/K\leq \alpha \\
\left(\frac{\alpha}{1-\alpha}\right) \left(\frac{1-u/K}{u/K}\right) & u/K\geq \alpha,
\end{array}\right.
\end{equation}
in which the first piece of $u f(u)$ connects the unstable zero at $u=0$ to the maximum at
$u=\alpha K$, while the second piece joins the maximum with the stable
zero at $u=K$.  The parameter $0\leq\alpha\leq1$ in (\ref{eq:linear}) represents the
relative position of the maximum of the nonlinearity between the stable and
unstable zeros of the system. If $\alpha=1/2$, the derivatives of
the logistic nonlinearity and its piecewise representation become equal at $u=0$ and $u=K$. Since the tail of a front solution is determined by the limit (\ref{eq:taillimitsolution}), the  choice $\alpha=1/2$ ensures that the spatial dependence of the tail (shoulder) front in the logistic nonlinearity
coincides with the tail (shoulder) front when $f(u)$ is given by (\ref{eq:linear}).
In the analytic treatment below, for the sake of  generality, we keep the parameter $\alpha$ unspecified.

Following the method of Ref. \cite{manne} it is straightforward to determine
the exact form of the traveling wave solution of Eq. (\ref{eq:driftdiffusion}) with $f$ given by \eqref{eq:linear}. Such a solution, denoted as $U(z,\gamma)$, where $\gamma=v_f/\beta$ and $v_f=2 \sqrt{D a}$ is given by
\begin{subequations} \label{eq:pwtwsolution}
\begin{align}
  U(az/\beta\leq0,\gamma)&=K-K(1-\alpha)\, \exp\left(\frac{az}{\beta}/\lambda_{1}\right),\\
  U(az/\beta>0,\gamma)&= B_{2}\,
  \exp\left(-\frac{az}{\beta}/\lambda_{2}\right) + B_{3}\,
  \exp\left(-\frac{az}{\beta}/\lambda_{3}\right), \label{eq:sumexp}
\end{align}
with
\begin{align}
\lambda_{1} &= (2/\gamma^2)
  \left(\sqrt{1+\frac{\alpha}{1-\alpha}\gamma^2}-1\right), &
  B_{2}       &= (1/2) K \left(\alpha-\xi\right),\\
  \lambda_{2} &= (2/\gamma^2) \left(1+\sqrt{1-\gamma^2}\right), &
  B_{3}           &=  (1/2) K \left(\alpha+\xi\right),\\
  \lambda_{3}&= (2/\gamma^2) \left(1-\sqrt{1-\gamma^2}\right), &
  \xi&=-(1-\alpha)\frac{\sqrt{1+\frac{\alpha}{1-\alpha}\gamma^2}}{\sqrt{1-\gamma^2}}+\frac{1}{\sqrt{1-\gamma^2}}.
\end{align}
\end{subequations}
By requiring that $u$ is always positive or
zero, a condition
for \eqref{eq:pwtwsolution} emerges as
\begin{equation} \label{eq:twcondition}
|\gamma|\leq 1.
\end{equation}
This bound represents
a constraint between the Fisher speed $v_f$ and the relative (to the medium) speed of the front solution $\beta$. Equation (\ref{eq:twcondition}) is equivalent to the minimum speed requirement for traveling wave of reaction-diffusion systems and it expresses the fact that the front speed must be faster or equal to the diffusive speed.
\begin{figure}[!htbp]
  \includegraphics[width=\columnwidth]{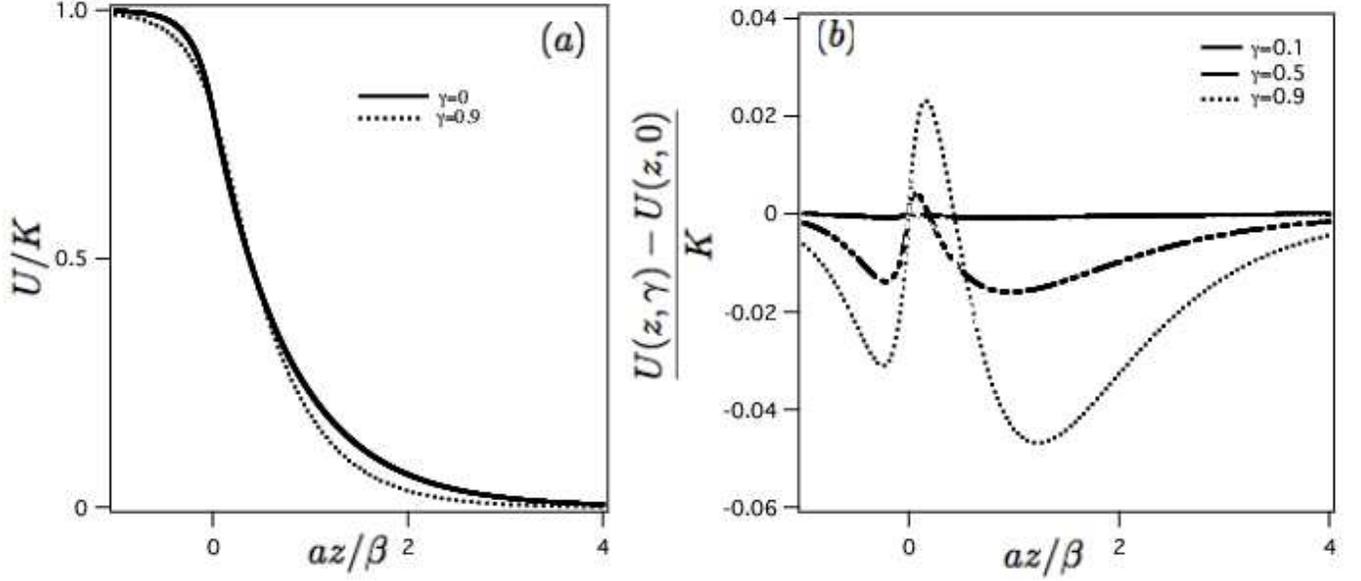}
\caption{Comparison of the shape of the traveling wave solution (\ref{eq:pwtwsolution}) with and without diffusion. On the left, in (a), a front solution with $\gamma=0$ and one with $\gamma=0.9$ are depicted.
On the right, in (b), $\Delta U(z,\gamma)=U(z,\gamma)- U(z,0)$ between Eq. (\ref{eq:pwtwsolution}) with $\gamma=0.1,0.5,0.9$ and the case with $\gamma=0$ are shown.}
\label{fig:pwdiff}
\end{figure}

\begin{figure}[!htbp]
  \includegraphics[width=0.6\columnwidth]{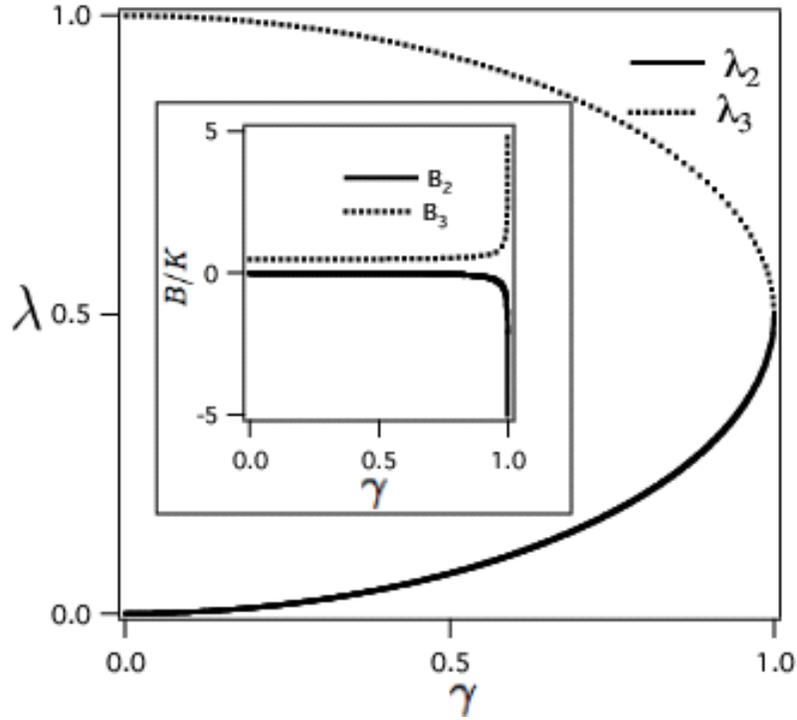}
\caption{Plot of the dependence on $\gamma$ of the parameters associated with the traveling front solution (\ref{eq:sumexp}). The characteristic lengths of the exponentials $\lambda_2$ and $\lambda_3$ are depicted between the values $\gamma=0$ and $\gamma=1$. In the inset the multiplicative factors $B_{2}$ and $B_{3}$ are also plotted.}
\label{fig:pltwdisplay}
\end{figure}

The parameter $\gamma$ is a measure of the relative diffusion in the system. In Fig. (\ref{fig:pwdiff}) we show how the front solution changes as diffusion increases. As in reaction diffusion equations, it is clear in Fig. (\ref{fig:pwdiff}b) that, as diffusion increases, the tail of $U(z,\gamma)$ acquires a shallower profile. The steepness of the tail as function of $\gamma$ can be determined from
Eq. (\ref{eq:sumexp}). The front tails are sums of two exponentials each one with characteristic lengths ($\lambda_{2}$ and $\lambda_{3}$) and multiplicative weighing factors ($B_{2}$ and $B_{3}$), all functions of $\gamma$. When $\gamma=0$, $B_{2}=0$ and $\lambda_{2}=0$. Eq. (\ref{eq:sumexp}) then contains \textit{only} one exponential term. The tail of the front solution is thus given by
\begin{equation} \label{eq:gamma0}
U\left(az/\beta>0,0\right)=B_3 \exp\left(-\frac{az}{\beta
\lambda_3}\right)=\alpha\,K\, \exp\left(-a z/\beta\right).
\end{equation}
At the other extreme, i.e., when $\gamma=1$, the characteristic lengths $\lambda_2$ and $\lambda_3$ are equal to each other, the divergent part of the coefficients, $\xi$, cancels in the sum, and the tail of the front solution becomes
\begin{equation} \label{eq:gamma1}
U\left(az/\beta>0,1\right)=(B_2+B_3) \exp\left(-\frac{az}{\beta 2}\right)+
=\alpha\, K\, \exp\left(-\frac{1}{2}a z/\beta\right).
\end{equation}
For values $0<\gamma<1$, both exponentials contribute to the tail shape but for small $\gamma$ values
$B_{2}$ is negligible and $\lambda_{3}\gg \lambda_{2}$ therefore making the second exponential in (\ref{eq:sumexp}) dominate. This dependence is plotted in Fig. \ref{fig:pltwdisplay}.

The use of the piecewise representation has allowed us to conduct an exact analysis above. When such a representation is not used, a perturbation scheme must be employed. The behavior shown in Fig. \ref{fig:pltwdisplay} lends support to the idea that a quantity directly related to $\gamma$
should serve as the small parameter on which to base the development of a perturbation technique for traveling front solutions.
We develop such a scheme in the next subsection.

\subsection{Perturbation treatment for traveling fronts}
The analysis presented here is a generalization of the technique proposed by Canosa
\cite{canosa1973} for studying traveling fronts of the Fisher equation.

In the presence of diffusion, the traveling fronts of Eq. (\ref{eq:driftdiffusion})
are solutions of the following equation
\begin{equation}
\frac{\gamma^2}{4} U_{\zeta\zeta}+ U_{\zeta} + U f(U)=0,
\end{equation}
where $\zeta=az/\beta$ and $z=x-ct$ where $c$ is the
absolute speed of the traveling front.

The quantity $\xi=(\gamma/2)^2 \le1/4$
is the perturbation parameter we will choose. A standard expansion of
the traveling front in powers of $\xi$ in Eq. (\ref{eq:driftdiffusion})
gives the identity
\begin{equation}
U_0'(\zeta) + \sum_{n=1}^\infty \xi^n \left( U_n'(\zeta) + U_{n-1}''(\zeta) \right)%
+F\left[U_0+\sum_{n=1}^\infty \xi^n U_n\right]=0,
\end{equation}
where $F(u)=uf(u)$. By Taylor-expanding the function $F(u)$, and rearranging each term in powers of $\xi$, it is possible to find the approximation of order $n$ in $\xi$ to the traveling front solution of Eq. (\ref{eq:driftdiffusion}). The zeroth order of such an expansion is given by the solution of
\begin{equation}
\frac{d U_0}{d \zeta} + F(U_0) = 0 \label{eq:zerothorder}
\end{equation}
which has been analyzed in the previous sections.
The next order is given by
\begin{equation}
\frac{d U_1}{d \zeta} + \frac{d^2 U_0}{d \zeta^2} + U_1 F'(U_0) = 0,
\end{equation}
whose solution is \cite{murray}
\begin{equation}
U_1(\zeta)=-U_0'(\zeta) \ln\left(4\, |U_0'(\zeta)|\right)
\end{equation}
The next order, the second, is given by
\begin{equation}
\frac{d U_2}{d \zeta} + U_2 F'(U_0) + \frac{d^2 U_1}{d \zeta^2} + \frac{1}{2!}F''(U_0) U_1^2 =0
\end{equation}
which is a linear equation for $U_2$, and all the other functions are given by solutions
of the previous orders.

To arbitrary order $n$, we have
\begin{equation}
\frac{d U_n}{d \zeta} + \frac{d^2 U_{n-1}}{d \zeta^2} + U_n F'(U_0) + h(\{U_{i<n}\}) = 0
\end{equation}
where $h(\{U_{i<n}\}$ are the rest of the terms that come from the $n$th power of $\xi$.

Canosa\cite{canosa1973} and Murray\cite{murray} have provided up to to first order for the Fisher case.
For the trigonometric nonlinearity with $f(u)$ given by (\ref{eq:lgps}) studied earlier in the present paper,
application of this perturbation techniques gives for the first order correction
\begin{equation}
U_1(z)= \frac{1}{\pi} \textrm{sech}(a z/\beta) \ln\left(\textrm{sech}(a z/\beta)\right).
\end{equation}

\begin{figure}[!htbp]
  \centering
  \includegraphics[width=\columnwidth]{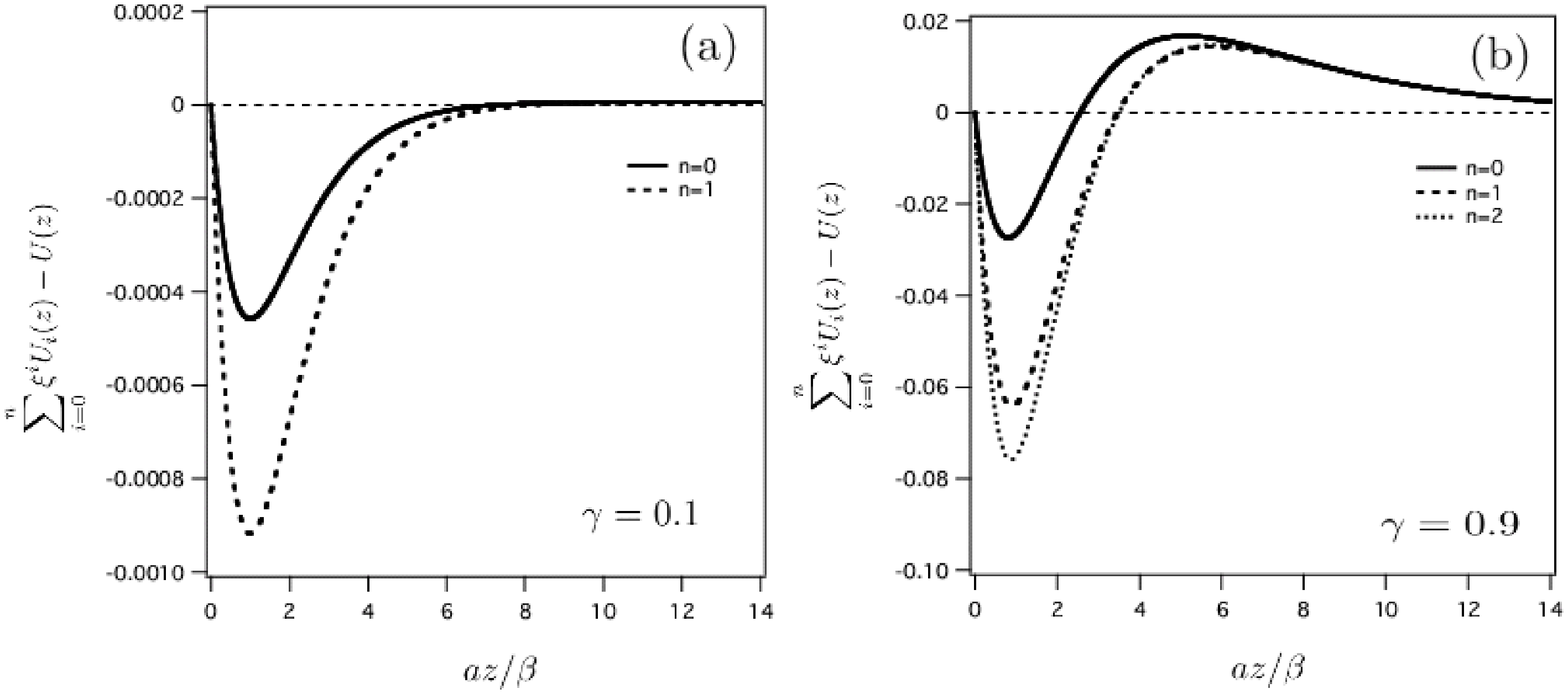}
  \caption{Difference between exact solution of the piecewise linear function and its zeroth, first and second order approximations in our perturbative treatment  for two different values: $\gamma=0.1$ and $\gamma=0.9$ in (a) and (b) respectively. Notice that, although not shown in the graph, as the order of the perturbation increases the curve difference converges rapidly to a curve which represents the error upper bound.}
  \label{fig:exactandnosoexact}
\end{figure}

To further analyze the validity of the perturbation scheme,
 we use the exact expression for the traveling fronts of the piecewise
linear function and compare it with the results obtained from the zeroth,
first, and second order perturbation.
Fig. \ref{fig:exactandnosoexact} shows the differences between the solutions
for two different values of the diffusion constant represented by $\gamma=0.1$ and $\gamma=0.9$.
We see that, for small $a z/\beta$, the lower orders are \emph{more} accurate than the higher orders.
In space this corresponds to that part of the front profile close to the value where $U(z)=1/2$. However, for the rest of the $az/\beta$ range, the higher orders in the perturbation get closer to the actual solution. It is worth noticing that, for $\gamma=0.9$, which represents a high value of diffusion, the error committed is still low.
Perhaps unexpectedly, for low values of $az/\beta$, the lower orders are more accurate.
The convergence of the orders is fast: for $\gamma=0.1$ the second order is just overlaped by the first order.

The above analysis has uncovered a perhaps obvious but important point
regarding the perturbation parameter. In \eqref{eq:driftdiffusion}, there are two
characteristic velocities: the Fisher velocity $v_f = \sqrt{Da}$ and
the medium velocity $v$. It may be tempting to suppose that the
appropriate perturbation parameter is their ratio \cite{gk}. This is
not correct because the entire effect of the medium velocity $v$ can
be eliminated by a transformation of the frame of reference.
Our analysis above shows that it is crucial to take into consideration
the third velocity that exists in the traveling wave context in this
problem, viz. the velocity of the traveling wave $c$. The appropriate
perturbation parameter when we consider a traveling wave is
$v_f/(c-v)$ and our conclusion is that for diffusion to be
considered as a perturbation on the pure convective analysis that we
have presented in this paper, it is necessary that the velocity of the
traveling wave exceed the medium velocity by an amount larger than the
Fisher velocity.

\section{Summary} \label{sec:conclusions}
Several physically realizable situations exist in various areas of
nonlinear population dynamics wherein the behavior is controlled
primarily by the nonlinearities and the convective element of the
motion, the diffusive contribution to the motion being small. The
typical example is the behavior of bacteria in a petri dish
\cite{nelson,ana,gk}, where the so-called "wind" motion 
induced by moving masks is made
stronger than the inherent diffusive motion. A study of such situations
has been provided in the present paper by first analyzing the case for
no diffusion (sections \ref{sec:application},\ref{sec:travelfront}) and then extending the analysis to small
diffusion (section \ref{sec:diffusion}). Explicit results have been obtained for a
variety of nonlinearities on the basis of a straightforward prescription
\cite{k}, for the full initial value problem. Traveling waves
have been analyzed explicitly for several cases. Interesting
consequences such as the "pyramid effect" for sinusoidal nonlinearities
have been shown to occur. The effect of small diffusion has been
incorporated through exact analysis for piecewise representation of the
nonlinearities on the one hand and through perturbative calculations on
the other. 

\begin{acknowledgments}
This work was supported in part by DARPA under grant no. DARPA-N00014-03-1-0900, by NSF/NIH Ecology of Infectious Diseases under grant no. EF-0326757, and by the NSF under grant no. INT-0336343.
\end{acknowledgments}


\begin{thebibliography}{99} 
\bibitem{fisher1937} R.A. Fisher, Ann. Eugen. \textbf{7}, 355 (1937).
\bibitem{murray} J.D. Murray, \textit{Mathematical Biology}, 2nd
  edition (Springer, New York, 1993).
\bibitem{skellam} J.C. Skellam. Biometrika \textbf{38}, 196 (1951)
\bibitem{kk} V.M. Kenkre and M.N. Kuperman, Phys. Rev. E \textbf{67},
  051921 (2003).
\bibitem{bkk} M. Ballard, V.M. Kenkre, and M.N. Kuperman, Phys. Rev. E
  \textbf{70}, 031912 (2004)
\bibitem{ak} G. Abramson and V.M. Kenkre, Phys. Rev. E \textbf{66},
  011912 (2003).
\bibitem{akyp} G. Abramson, V.M. Kenkre, T. Yates, and R.R. Parmenter,
  Bull. Math. Bio. \textbf{65}, 519 (2003).
\bibitem{kpasi} V.M. Kenkre, in \textit{Modern Challenges in
    Statistical Mechanics: Patterns, Noise, and the Interplay of
    Nonlinearity and Complexity}, V. M. Kenkre and K. Lindenberg,
  eds., AIP Proc. vol. \textbf{658} (2003), p. 63.
\bibitem{manne} K.K. Manne, A.J. Hurd, and V.M. Kenkre, Phys. Rev. E
  \textbf{61}, 4177 (2000).
\bibitem{abk} G. Abramson, A.R. Bishop, and V.M. Kenkre, Phys. Rev. E
  \textbf{64}, 66615 (2001).
\bibitem{fkk} M.A. Fuentes, M.N. Kuperman, and V.M. Kenkre, Phys. Rev.
  Lett. \textbf{91}, 158104 (2003).
\bibitem{fk} M.A. Fuentes, M. Kuperman, and V.M. Kenkre, J. Phys. Chem. B \textbf{108}, 10505 (2004).
\bibitem{nelson} D.R. Nelson and N.M. Shnerb, Phys. Rev. E \textbf{58},
  1383 (1998); K.A. Dahmen, D.R. Nelson, and N.M. Shnerb, J. Math. Biology
  \textbf{41}, 1 (2000).
\bibitem{ana} A.L. Lin, B.A. Mann, G. Torres-Oviedo, B. Lincoln, J.
  Kas, and H.L. Swinney, Biophys. J. \textbf{87}, 75 (2004).
\bibitem{gk} L. Giuggioli and V.M. Kenkre, Physica D \textbf{183}, 245
  (2003).
\bibitem{lgthesis} L. Giuggioli, Ph.D. Thesis (University of New
  Mexico).
\bibitem{k} V.M. Kenkre, Physica A \textbf{342}, 242 (2004).
\bibitem{allee} W.C. Allee,  \textit{The Social Life of Animals}, (Beacon Press, Boston,1938).
\bibitem{courchamp} F. Courchamp, T. Clutton-Brock, and B. Grenfell, Trends Ecol. Evol. \textbf{14}, 405 (1999).
\bibitem{SS} P.A. Stephens and W.J. Sunderland, Trends Ecol. Evol. \textbf{14}, 401 (1999).
\bibitem{fowlerbaker} C.W. Fowler and J.D. Baker, Rep. Int. Whaling Comm. \textbf{41}, 545 (1999).
\bibitem{rosas2002} A. Rosas, C.P. Ferreira, and J.F. Fontanari, Phys.
  Rev. Lett. \textbf{89}, 188101 (2001).
\bibitem{soli} M. Remoissenet, \textit{Waves Called Solitons: Concepts and Experiments}, 3rd edition, (Springer, Berliln, 1999).
\bibitem{hypergeometric} Ed. M. Abramowitz and I.A. Stegun, \textit{Handbook of Mathematical Functions}, (Dover, New York, 1970)
\bibitem{char} I.G. Petrovsky, \textit{Lectures on Partial Differential Equations}, (Dover, New York, 1991)
\bibitem{horacio} G. Izus, R. Deza, C. Borzi, and H.S. Wio, Physica A \textbf{237}, 135 (1997).
\bibitem{canosa1973} J. Canosa, IBM. J. Res. and Dev. \textbf{17} 307 (1973).

\end{thebibliography}
\end{document}